\documentstyle[12pt]{article}
\begin{document}

\begin{center}
{\bf{A q-Generalization of Product Densities and Janossy Function in Stochastic Point Processes}}

\vspace{0.5cm}

R.Parthasarathy and R.Sridhar {\footnote{e-mail addresses: sarathy, sridhar@imsc.res.in }} \\
The Institute of Mathematical Sciences \\
C.P.T.Campus, Taramani Post \\
Chennai 600 113, India. \\
\end{center}

\vspace{0.5cm}

{\noindent{\it{Abstract}}}

\vspace{0.5cm}

A q-generalization of the product densities in stochastic point processes is developed. The properties of these functions
are studied and a q-generalization of the usual $C^r_s$ coefficients is obtained. This for fixed q-number of particles
coincides with the q-Stirling numbers of the second kind. The q-product densities are investigated using q-Poisson
distribution and this shows that the stochastic point processes involving consistent q-generalization are inherently
correlated. A closely related function to q-product densities is a q-generalized Janossy function and a relation between the
two is established.          

\newpage

{\noindent{\bf{I.Introduction}}}

\vspace{0.5cm}

In the investigation of the cosmic ray cascades, as applications of the theory of stochastic processes [2], one deals with a
stochastic variable representing the number of particles distributed in a continuous infinity of states characterized by a
parameter $E$ (the energy for example). In such a study,  
Ramakrishnan [12] in 1950 introduced the concept of {\it{Product
Densities}} which played a crucial role in studying a statistical assembly of particles distributed over a continuous parameter. In
cosmic ray cascades, the number of particles of a given energy $E$ or after traversing a thickness $t$ is considered to be a
{\it{stochastic variable}}. The energy $E$ or the thickness $t$ is taken to be a continuous variable. Thus, a stochastic
variable $N$ taking discrete values, depends on a continuous parameter $E$ or $t$. 
Such processes are called {\it{Stochastic Point Processes}} since the continuous variable $E$
or $t$ for specific values represents a point in the energy or thickness axis.  
The use of {\it{Product Densities}} greatly
simplified the treatment of the stochastic variable $N$. 
Detailed account of the applications of product densities in various stochastic processes
ranging from cosmic ray cascades to biology  can be
found in the review of Ramakrishnan [13], and the books by Srinivasan [14], Bartlett [1] and Bharucha-Reid
[3]. 

\vspace{0.5cm}

The moments of the distribution of the number of particles (stochastic variable) over specified intervals $(E, E+\triangle E)$
can be obtained from product densities. The remarkable achievement of Ramakrishnan [12] is the investigation of the  higher
order moments in terms of product densities of higher degree $(>2)$. This led him to introduce the $C^r_s$ coefficients which can
be identified with Stirling numbers of the second kind, for fixed $N$, thereby bringing a connection to a 'number-theoretic' result
of expressing $m^{th}$ power of a variable in terms of lower powers using falling factorials. The sum of $C^r_s\ (s\leq r)$ over
all posible values of $s$ is the Bell's number $B^{(r)}$. 

\vspace{0.5cm}

It is the purpose of this paper to consider a q-generalization of the product densities by addressing a stochastic point
process for q-numbers [4] distributed over a continuos variable. Such q-numbers will be called
q-stochastic variable. 
This is carried out by studying the statistical properties of a
q-stochastic variable (q-number) with a related q-extension of the
product densities.   
The continuous variable $E$ or $t$
upon which the q-stochastic variable depends, is taken to be an ordinary variable. This consideration of studying the statistical properties of q-stochastic variable 
distributed over continuous variable $E$, can be used to describe light wave as a stochastic
variable, namely, the possible field realizations are stochastic. As the fields are distributed
over the space, the q-stochastic variable approach which {\it{sans}} the integer values, is apt
to describe laser light as well as thermal light [8].   

\vspace{0.5cm}

The precise definition of q-product densities will be given in section.III.   
During the course of the study,  
 we introduce a q-generalization of the $C^r_s$ coefficients which become q-Stirling numbers of the second kind when
the q-number of particles distributed, is fixed.    
One of the results of this classical analysis is a derivation of q-Bell
number (See section.IV). In the literature, q-Bell number is
investigated using q-Boson coherent states by Katriel [7]. As q-Stirling
numbers of the second kind and the q-Bell number are the results of a
combinatorial analysis of the q-numbers, it will be appropriate to
obtain them using classical stochastic process as done here, rather than
the use of semi-classical approach in [7]. Further, in this present
study,  
q-product densities are investigated using probability distribution function and a q-generalization
of the Poisson distribution is illustrated. {\it{We find a natural role for the q-Poisson distribution}}. 

\vspace{0.5cm}

The concept of product densities in stochastic processes is very briefly reviewed in section.II.   
In section.III, the q-product densties are introduced and their properties derived. This leads to the
q-analogue of the $C^r_s$ coefficients which for a given number of particles coincides with q-Stirling numbers of the second
kind. They are evaluated for a few representative cases. A recursion formula for $q-C^r_s$ coefficients is derived. 
q-Bell number is
discussed in Section.IV. The use of q-Poisson distribution in the context of q-product densities is studied in section.V and
section VI is devoted to q-Janossy function and its relation to q-product density function. The results are summarized in
section.VII with discussion. 
   
\vspace{0.5cm}

{\noindent{\bf{II.Brief Review of Product Density}}}

\vspace{0.5cm}

Let $N(E(t))$ be a stochastic variable (number of particles) with parametric values $\leq E$ at $t$. We shall suppress $t$
hereafter. Then $dN(E)$ is the number of particles in the range $E$ and $E+dE$. Ramakrishnan [12] assumed that the probability that
there occurs one particle in $dE$ is proportional to $dE$, while that for the occurence of $n$ particles in $dE$ is proportional to
$(dE)^n$. The average number of particles ${\cal{E}}(dN(E))$ in $dE$  
is represented by a function $f_1(E)$ (product density of degree one)  
such that
\begin{eqnarray}
{\cal{E}}(dN(E))&=& f_1(E)dE.
\end{eqnarray}
Denoting the probability that $n$ particles occur in $dE$ by $P(n)$, we have 
\begin{eqnarray}
P(1)&=&f_1(E)dE + O((dE)^2), \nonumber \\
P(0)&=& 1-f_1(E)dE - O((dE)^2), \nonumber \\
P(n) &=& O((dE)^n)\ \ \ \ ;\ n>1.
\end{eqnarray}

The $r^{th}$ moment is given by
\begin{eqnarray}
{\cal{E}}\{n^r\}&=&\sum_{n} n^rP(n), \nonumber \\
                &=&{\cal{E}}\{ dN(E)\},
\end{eqnarray}
where the last step follows from (2).    

\vspace{0.5cm}

The expectation value of the product of the stochastic variables $dN(E_1)$ and $dN(E_2)$ 
has been defined in  
[12] as 
\begin{eqnarray}
{\cal{E}}\{dN(E_1)dN(E_2)\}&=& f_2(E_1,E_2)dE_1dE_2.
\end{eqnarray}
$f_2(E_1,E_2)$ is the product density of degree 2 and the right hand side of (4) is also the joint probability that a particle lies
in $dE_1$ and another in $dE_2$ when {\it{ $dE_1$ and $dE_2$ do not overlap, irrespective of the number of particles in other
ranges}}. When the intervals overlap, a degeneracy occurs and so in general,
\begin{eqnarray}
\int_{E_\ell}^{E_u}\ \int_{E_{\ell}}^{E_u} {\cal{E}}[dN(E_1)dN(E_2)]&=&\int_{E_{\ell}}^{E_u}f_1(E)dE \nonumber \\  
&+& \int_{E_{\ell}}^{E_{u}}
\int_{E_{\ell}}^{E_u}f_2(E_1,E_2)dE_1dE_2.
\end{eqnarray}
Product densities of higher degree are defined by 
\begin{eqnarray}
f_n(E_1,E_2,\cdots E_n)dE_1dE_2\cdots dE_n&=& {\cal{E}}\{dN(E_1)dN(E_2)\cdots dN(E_n)\},
\end{eqnarray}
denoting the joint probability that there lies one particle in $dE_1$, one in $dE_2$ and so on where the intervals $dE_1,
dE_2,\cdots dE_n$ do not overlap, irrespective of the number of particles in other ranges.

\vspace{0.5cm}

The case of $N$ particles distributed continuously in the $E$-space is of particular interest as then $N$ is a constant. 
Nevertheless, the distribution in $E$-space follows probability laws. We have in
this case,
\begin{eqnarray}
f_1(E)dE &=& Nf_1^0(E)dE, \nonumber \\
\int_{(whole\ range)} f_1^0(E)dE &=& 1,
\end{eqnarray}
where $f_1^0(E)dE$ represents the probability that any one of the $N$ particles selected at random lies in $dE$. It follows then 
\begin{eqnarray}
f_2(E_1,E_2)dE_1dE_2&=&N(N-1)\ f_!^0(E_1)f_1^0(E_2)dE_1dE_2, \nonumber \\
f_m(E_1,\cdots E_m)dE_1\cdots dE_m&=&N(N-1)\cdots (N-(m-1))f_1^0(E_1)\cdots f_1^0(E_m) \nonumber \\ 
 & & dE_1dE_2 \cdots dE_m,
\nonumber \\
f_N(E_1,\cdots E_N)dE_1\cdots dE_N&=&N!\ f_1^0(E_1)f_1^0(E_2)\cdots f_1^0(E_N)dE_1\cdots dE_N.
\end{eqnarray}         

\vspace{0.5cm}

The $r^{th}$ moment of the number of particles in any finite range $\triangle E=E_u-E_{\ell},\ {\cal{E}}(N^r_{\triangle E})$ can be
represented, after taking into account the degeneracies in the intervals $\{dE_i\}$, by
\begin{eqnarray}
{\cal{E}}(N^r_{\triangle E})&=&\sum_{s=1}^{r} C^r_s\int_{E_{\ell}}^{E_u}\cdots \int_{E_{\ell}}^{E_u}f_s(E_1,\cdots E_s)dE_1\cdots
dE_s,
\end{eqnarray}
where the coefficients $C^r_s$ are functions of $r$ and $s$ and do not depend on the function $f$. Applying (9) to the case in which
the number of particles $N$ is fixed, using (8) and integrating over the whole range, we get
\begin{eqnarray}
N^r &=& \sum_{s=1}^{r} C^r_s \frac{N!}{(N-s)!},
\end{eqnarray}
which identifies $C^r_s$ with the Stirling number of the second kind. It is to be noted that (9) is a general expression for the
$r^{th}$ moment of $N$ in terms of integrals of product densities of order $\leq r$ and the $C^r_s$ coefficients in (9) become
Stirling numbers of the second kind only when $N$ is fixed.     

\vspace{0.5cm}

Considering the case when there is no correlation between particles in any two different energy ranges, the product density of
degree $r$ is factored as 
\begin{eqnarray}
f_r(E_1\cdots E_r)&=& f_1(E_1)\cdots f_1(E_r),
\end{eqnarray}
and (9) becomes
\begin{eqnarray}
{\cal{E}}(N^r_{\triangle E})&=&\sum_{s=1}^{r} C^r_s {\{ {\cal{E}}(N_{\triangle E})\} }^s,
\end{eqnarray}
where the integrals are taken over the finite range $\triangle E$. It is to be noted that (12) coincides with the moments of a
variable with a Poisson distribution. To see this, let $P(n,\triangle E)$ be a distribution function of the number of particles $n$
in $\triangle E$ and $G(u,\triangle E)$ be its generating function, i.e.,
\begin{eqnarray}
G(u,\triangle E) &=& \sum_{n} u^n P(n,\triangle E).
\end{eqnarray}
The $r^{th}$ moment of $n$ is 
\begin{eqnarray}
{\cal{E}}(n^r_{\triangle E}) &=& \sum_{n} n^r P(n,\triangle E), \nonumber \\
&=& \Bigl\{ {\Big( u\frac{d}{du}\Big)}^r G(u,\triangle E)\Bigr\}_{u=1},
\end{eqnarray}
where the second line follows from (13). Expanding and regrouping this we have,
\begin{eqnarray}
{\cal{E}}(n^r_{\triangle E})&=&\sum_{s=1}^{r} b^r_s \Bigl\{ G^{(s)}(u,\triangle E)\Bigr\}_{u=1},
\end{eqnarray}
where $G^{(s)}(u,\triangle E)$ stands for the $s^{th}$ partial derivative of $G(u,\triangle E)$ with respect to $u$. For a Poisson
distribution,
\begin{eqnarray}
P(n,\triangle E) &=& e^{-{\cal{E}}(n_{\triangle E})} \frac{ ({\cal{E}}(n_{\triangle E}))^n}{n!}, \nonumber 
\end{eqnarray}
we have 
\begin{eqnarray}
G(u,\triangle E)&=& e^{(u-1){\cal{E}}(n_{\triangle E})}.
\end{eqnarray}    
Then (15) becomes 
\begin{eqnarray}
{\cal{E}}(n^r_{\triangle E})&=& \sum_{s=1}^{r} b^r_s {\Bigl\{ {\cal{E}}(n_{\triangle E})\Bigr\}}^s.
\end{eqnarray}
Comparing this with (12), we find $b^r_s=C^r_s$. Thus the case of no correlation between particles in any two different energy
ranges is governed by Poisson distribution. Although the result $b^r_s=C^r_s$ has been shown for Poisson distribution, since
$C^r_s$ coefficients do not depend on $f_s$, (15) holdsgood generally with $b^r_s$ replaced by $C^r_s$. 

\vspace{0.5cm}

{\noindent{\bf{III.q-generalization of Product Densities}}}

\vspace{0.5cm}

We shall now extend the theory of product densities to a slightly different type of stochastic problem, namely,  a q-extension
of the stochastic variable $N$ while keeping the continuous parameter $E$ as ordinary. Then the statistical properties of a q-stochastic variable $n$ taking
discrete values  
\begin{eqnarray}
{[n]} &=& \frac{1-q^n}{1-q},
\end{eqnarray}
depends on the continuous parameter $E$ and so we have a q-stochastic point process. When a q-stochastic variable takes value $[n(E)]$, we need to find
the q-number in the range $E$ and $E+dE$. 
The q-number of particles in the range $E$
and $E+dE$ {\it{is defined to be}} $([n(E+dE)-n(E)])$ {\it{which
is just $[dn(E)]$}}. 

\vspace{0.5cm}

Now we extend the considerations in the previous section, namely, the
probability that there occurs one particle in $dE$ is proportional to
$dE$ and this is consistent with our definition given above since
$[1]=1$. The probability for the occurrence of $[n]$ particles in $dE$
is taken to be proportional to $(dE)^n$. Then the average number of
particles in the interval $dE$, denoted by ${\cal{E}}([dn(E)])$, is
represented by a function $f^{(q)}_1(E)$ such that 
\begin{eqnarray}
{\cal{E}}([dn(E)])&=& f^{(q)}_1(E)\ dE. \nonumber 
\end{eqnarray}
Denoting the probability that $[n]$ particles in the interval $dE$ by
$P_q([n])$, we postulate (q-analogue of (2)), 
\begin{eqnarray}
P_q([1])=P_q(1) &=& f^{(q)}_1(E)dE + O((dE)^2), \nonumber \\
P_q([0])=P_q(0) &=& 1-f^{(q)}_1(E)dE - O((dE)^2), \nonumber \\
P_q([n]) &=& O((dE)^n)\ ;\ \ \ n>1. \nonumber 
\end{eqnarray}         

The $r^{th}$ moment in (3) gets generalized to 
\begin{eqnarray}
{\cal{E}}\{ [n]^r\} &=& \sum_{n} [n]^rP_q([n]), \nonumber \\
 &=& {\cal{E}}\{ [dn(E)]\}.
\end{eqnarray}
The last step in the above equation follows from the postulate stated
above. Since ${\cal{E}}([dn(E)])$ is proportional to $dE$, we are able
to maintain the result of [11] that all the moments of the q-stochastic
variable $[dn(E)]$ are equal to the probability that the stochastic
vriable assumes the value $[1]=1$. 
  
\vspace{0.5cm}

We shall denote the q-product densities by superscript $q$, \\ as $f_1^{(q)}(E), f_2^{(q)}(E_1,E_2),\cdots f_n^{(q)}(E_1,E_2,\cdots
E_n)$. The q-product densities are necessary for the following reasons. In attempting the q-extension, one 
considers a distribution of $[N]$ particles in the energy axis.  

That is, $[N]$ number of particles to be distributed in the intervals $dE_1,
dE_2, \cdots dE_N$ with $[1]=1$ particle in each energy interval. 
This can be done in $^{[N]}C_{[1]}=[N]!/([1]![N-1]!)=[N]$ number of ways, where $^{[n]}C_{[1]}$
is the q-Binomial coefficient,  
for the first interval $dE_1$. The remaining particles are  
$[N]-1$ and also this is the number of ways of putting in the second interval. So, for product density of degree 2, the
joint probability of putting $[1]$ particle each in $dE_1$ and $dE_2$, will be proportional to $[N]([N]-1)$ with the
{\it{same}} $f_2$ function as in (8). Proceeding in this way, with the {\it{same}} $f$ function
as in (8), we encounter $[N]([N]-1)([N]-2)\cdots ([N]-N+1)$ factor for the $N^{th}$ interval.
This will not exhaust the total number of particles. Thus, in dealing with q-numbers to be
distributed, we need to introduce q-product densities such that the joint probability of putting
$[1]$ particle  each in $dE_1$ and $dE_2$ will be proportional to $[N][N-1]$, that for putting
$[1]$ particle each in $dE_1, dE_2$ and $dE_3$ proportional to $[N][N-1][N-2]$ and so on. This
will lead to the joint probability of putting $[1]$ particle each in $dE_1,\cdots dE_N$
proportional to $[N]!$, thereby exhausting the total number of $[N]$ particles.     
The use of these q-product densities greatly simplifies the treatment, as will be shown subsequently. 

\vspace{0.5cm}

As the continuous parameters $E_1, \cdots E_n$ are taken as usual variables, {\it{the integration over $dE_1\cdots dE_n$ will be
ordinary integrals.}} So, (5) and (6) become 
\begin{eqnarray}
\int_{E_{\ell}}^{E_u} \int_{E_{\ell}}^{E_u} {\cal{E}}\{ dN(E_1)dN(E_2)\}&=& \int_{E_{\ell}}^{E_u} f_1^{(q)}(E)dE \nonumber \\
 &+& \int_{E_{\ell}}^{E_U} \int_{E_{\ell}}^{E_u} f_2^{(q)}(E_1,E_2)dE_1dE_2, \nonumber \\ 
f_n^{(q)}(E_1,\cdots E_n)dE_1\cdots dE_n&=&{\cal{E}}\{dN(E_1)\cdots dN(E_n)\}.
\end{eqnarray}       

In proceeding further, the case of $[N]$ number of
particles distributed continuously in the $E$-space is of interest. Here $[N]$ is a constant. However, the distribution in
$E$-space follows probability laws which we are to find out. Therefore, we can then write in this case,
\begin{eqnarray}
f_1^{(q)}(E)dE &=& [N] f_1^{(q)0}(E) dE, \nonumber \\
\int_{(whole range)}f_1^{(q)0}(E)dE &=&  1, 
\end{eqnarray}
where $f_1^{(q)0}(E)dE$ represents the probability that any one ($[1]=1$) of the $[N]$ particles selected at random lies in $dE$.  
Then,
\begin{eqnarray}
f_2^{(q)}(E_1,E_2)dE_1dE_2&=&[N][N-1]f_1^{(q)0}(E_1)f_1^{(q)0}(E_2)dE_1dE_2, \nonumber \\
f_m^{(q)}(E_1,\cdots E_m)dE_1\cdots dE_m&=&\frac{[N]!}{[N-m]!}f_1^{(q)0}(E_1)\cdots f_1^{(q)0}(E_m)dE_1\cdots dE_m, \nonumber \\
f_N^{(q)}(E_1,\cdots E_N)dE_1\cdots dE_N&=&[N]!f_1^{(q)0}(E_1)\cdots f_1^{(q)0}(E_N)dE_1\cdots dE_N.
\end{eqnarray}      

\vspace{0.5cm}

The $r^{th}$ moment of the q-number of particles in any finite range $\triangle E=E_u-E_{\ell}$, namely ${\cal{E}}\{
[N]^r_{\triangle E}\}$ is represented by 
\begin{eqnarray}
{\cal{E}}\{ [N]^r_{\triangle E}\}&=&\sum_{s=1}^{r} {\cal{C}}^r_s\int_{E_{\ell}}^{E_u}\cdots
\int_{E_{\ell}}^{E_u}f_s^{(q)}(E_1,\cdots E_s)dE_1\cdots dE_s,
\end{eqnarray}
where ${\cal{C}}^r_s$ is the q-analogue of the $C^r_s$ coefficient in (9). From (23) and (22), using (21), for $[N]$ fixed, we
obtain,
\begin{eqnarray}
{[N]}^r &=& \sum_{s=1}^{r} {\cal{C}}^r_s \frac{[N]!}{[N-s]!},
\end{eqnarray}
when integrated over the whole range of $dE$. This expresses for $[N]^r$ in terms of 'falling q-factorials'. The coefficients
${\cal{C}}^r_s$ can be identified with the q-Stirling numbers of the second kind. By writing 
\begin{eqnarray}
{[N]}^{r+1} &=& \sum_{s=1}^{r+1} {\cal{C}}^{r+1}_s \frac{[N]!}{[N-s]!}, \nonumber 
\end{eqnarray}
and comparing this with the identity $[N]^{r+1}=[N]^r [N]$ and using (24), we obtain a recursion relation for ${\cal{C}}^r_s$ as 
\begin{eqnarray}
{\cal{C}}^{r+1}_s &=& q^{s-1} {\cal{C}}^r_{s-1} + [s] {\cal{C}}^r_s.
\end{eqnarray}   

\vspace{0.5cm}

Numerical values for ${\cal{C}}^r_s$ coefficients can be obtained from (24). For $r=1$, (24) gives ${\cal{C}}^1_1=1$. From the
definition of $[n]$ (18), it follows, $s<n$,
\begin{eqnarray}
{[n-s]} &=& \frac{[n]-[s]}{q^s}.
\end{eqnarray} 
For $r=2$, using (24) and (26) we find,
\begin{eqnarray}
{\cal{C}}^2_1\ =\ 1 &;& {\cal{C}}^2_2\ =\ q.
\end{eqnarray}
Similarly, for $r=3$, we find,
\begin{eqnarray}
{\cal{C}}^3_1=1, &{\cal{C}}^3_2=q([1]+[2])&, {\cal{C}}^3_3=q^3.
\end{eqnarray}
The values in (28) can also be obtained using the recursion relation (25) and (27) (as input) with the conditions ${\cal{C}}^r_s=0$,
if $s=0$ or $r<s$. Higher order ${\cal{C}}^r_s$ coefficients can be obtained recursively from (25).  

\vspace{0.5cm}

{\noindent{\bf{IV.q-Bell Number}}}

\vspace{0.5cm}

In the usual case described in Section.II, the Bell number is defined as
\begin{eqnarray}
B^{(r)}&=& \sum_{s=1}^r C^r_s,
\end{eqnarray}
and a series expansion for $B^{(r)}$ is given by Dobinsky's formula [11]
\begin{eqnarray}
B^{(r)} &=& \frac{1}{e}\sum_{n=1}^{\infty} \frac{n^r}{n!}.
\end{eqnarray}
In view of the $q-C^r_s$ coefficients ${\cal{C}}^r_s$, we can define the q-Bell number as 
\begin{eqnarray}
{\cal{B}}^{(r)} &=& \sum_{s=1}^{r} {\cal{C}}^r_s,
\end{eqnarray}
which gives  using (25)-(28), 
\begin{eqnarray}
{\cal{B}}^{(1)}&=&1, \nonumber \\ 
{\cal{B}}^{(2)}&=&[2], \nonumber \\  
{\cal{B}}^{(3)}&=&1+q\{ [1]+[2]\}+q^3, \nonumber \\  
{\cal{B}}^{(4)}&=&1+q\{[1]^2+[2]^2+[1][2]\}+q^3\{[1]+[2]+[3]\}+q^6,
\end{eqnarray}
and so on. A q-Dobinsky formula [9] can be obtained from (24).
Rewriting (24) as   
\begin{eqnarray}
\frac{[N]^r}{[N]!}&=&\sum_{s=1}^r {\cal{C}}^r_s \frac{1}{[N-s]!},
\end{eqnarray}
and multiplying by ${\lambda}^N$ and summing over $N$ from 1 to $\infty$, we find 
\begin{eqnarray}
\sum_{s=1}^r{\cal{C}}^r_s {\lambda}^s &=& \frac{1}{e_q(\lambda)}\sum_{N=1}^{\infty} \frac{[N]^r {\lambda}^N}{[N]!}.
\end{eqnarray}
Setting $\lambda =1$ we have 
\begin{eqnarray}
\sum_{s=1}^{r}{\cal{C}}^r_s\ =\ {\cal{B}}^{(r)}&=& \frac{1}{e_q(1)}\sum_{N=1}^{\infty} \frac{[N]^r}{[N]!}.
\end{eqnarray}
It follows from (35) that ${\cal{B}}^{(1)}=1$.
And ${\cal{B}}^{(2)}=\frac{1}{e_q(1)}\sum_{N=1}^{\infty} \frac{[N]^2}{[N]!}$. Writing $[N]^2=[N]+q[N][N-1]$, we realize
${\cal{B}}^{(2)}=[2]$. Similarly it can be verified that (32) is reproduced by (35). Although the q-Bell number and the
corresponding Dobinsky formula have been obtained by Katriel [7] using q-coherent states, our derivation here uses the
q-product densities and consitutes an independent approach. {\it{A rigorous derivation of the formula
(35) for the q-Bell number using q-Poisson distribution will be given at the end of Section.V}}.

\vspace{0.5cm}

{\noindent{\bf{V.q-Product densities and q-Poisson distribution}}}

\vspace{0.5cm}

So far, we have not used any specific distribution for q-product densities. We first recall a result in the theory of product
densities, namely, {\it{if $G(u,\triangle E)$ is the probability generating function to $P(n,\triangle E)$, the probability
distribution function of the total number of particles $n$ in the energy range $\triangle E$ and if $f_k(E_1,\cdots E_k)$ is the
product density of degree $k$, then }}
\begin{eqnarray}
\int_{E_1}\cdots \int_{E_k}f_k(E_1,\cdots E_k)dE_1\cdots dE_k&=&{\Bigl\{ \frac{{\partial}^kG(u,\triangle E)}{{\partial}u^k}
\Bigr\}}_{u=1},
\end{eqnarray}
{\it{where the integration with respect to each variable is over the finite range $\triangle E$.}}   
This can be derived by comparing (9) with (15) and using $C^r_s=b^r_s$.

\vspace{0.5cm}

Now, we wish a q-analogue of (36). In our earlier steps, we have considered q-number of particles $[n]$. From the definition of the
generating function $G(u,\triangle E)$ in (13), it follows that the variable $u$ is associated with $n$. 
Therefore when $n$ is replaced by the
q-number $[n]$, the variable $u$ should be treated consistently such as the differentiation with respect $u$ should be
q-differentiation now. Then the q-analogue of (36) will be 
\begin{eqnarray}
\int_{E_1}\cdots \int_{E_k}f^{(q)}_k(E_1,\cdots E_k)dE_1\cdots dE_k&=&{\Bigl\{ \frac{{\partial}^k_q{\cal{G}}(u,\triangle E)}
{{\partial}_qu^k}\Bigr\}}_{u=1},
\end{eqnarray}  
where the integrations over $E$ are ordinary integrals. The above q-analogue means {\it{if ${\cal{G}}(u,\triangle E)$ is the
probability generating function to $P_q([n],\triangle E)$, the probability distribution function of the q-number $[n]$ of particles
in the energy range $\triangle E$ and if $f_k^{(q)}(E_1,\cdots E_k)$ is the q-product density of degree $k$, then (37) relates the
two through q-differentiation.}} In order to prove (37) we proceed as below. The probability generating function
${\cal{G}}(u,\triangle E)$ is related to the probability distribution function $P_q([n],\triangle E)$ by 
\begin{eqnarray}
{\cal{G}}(u,\triangle E) &=& \sum_{n} u^n P_q([n],\triangle E).
\end{eqnarray}
The $r^{th}$ moment of $[n]$ is given by
\begin{eqnarray}
{\cal{E}}\{ [n]^r_{\triangle E}\} &=& \sum_{n} [n]^r P_q([n],\triangle E),
\end{eqnarray}
which can be written as
\begin{eqnarray}
{\cal{E}}\{[n]^r_{\triangle E}\}&=&{\Bigl\{ {\Big( u\frac{{\partial}_q}{{\partial}_qu}\Big)}^r {\cal{G}}(u,\triangle E)
\Bigl\}}_{u=1},
\end{eqnarray}
since 
\begin{eqnarray}
{\Big( u\frac{{\partial}_q}{{\partial}_qu}\Big)}^r{\cal{G}}(u,\triangle E)&=&\sum_{n}[n]^ru^rP_q([n],\triangle E), \nonumber 
\end{eqnarray}
using (38). The terms in (40) can be regrouped to give        
\begin{eqnarray}
{\cal{E}}\{[n]^r_{\triangle E}\}&=&\sum_{s=1}^{r}{\cal{B}}^r_s {\Bigl\{ {\cal{G}}^{(s)}(u,\triangle E)\Bigr\}}_{u=1},
\end{eqnarray}
where ${\cal{G}}^{(s)}(u,\triangle E)$ stands for the $s^{th}$ q-derivative with respect to $u$.  

\vspace{0.5cm}

Now using (38) in (37) and noting 
\begin{eqnarray}
\frac{{\partial}_q^k{\cal{G}}(u,\triangle E)}{{\partial}_qu^k}&=&\sum_{n}\frac{[n]!}{[n-k]!}P_q([n],\triangle E), \nonumber 
\end{eqnarray}
we have 
\begin{eqnarray}
\int_{E_1}\cdots \int_{E_k}f^{(q)}_k(E_1,\cdots E_k)dE_1\cdots dE_k&=&\sum_{n-k=0}^{\infty}
\frac{[n]!}{[n-k]!}P_q([n],\triangle E.)
\end{eqnarray}

\vspace{0.5cm}

The above relation holdsgood for any q-distribution. Now, we specialize in q-Poisson distribution. The generating function for
q-Poisson distribution is 
\begin{eqnarray}
{\cal{G}}_{Poisson}(u,\triangle E) &=& e_q^{u{\cal{E}}([n],\triangle E)}\ e_{\frac{1}{q}}^{-{\cal{E}}([n],\triangle E)},
\end{eqnarray}
where $e_{q}^x$ is the q-expponential function $\sum_{n=0}^{\infty}\frac{x^n}{[n]!}$ and $e_{\frac{1}{q}}^x$ is the second
q-exponential function with $q$ replaced by $\frac{1}{q}$ and we have $e_q^x\ e_{\frac{1}{q}}^{-x}=1$ [10]. Then (38) becomes
\begin{eqnarray}
e_{\frac{1}{q}}^{-{\cal{E}}([n],\triangle E)}\sum_{n=0}^{\infty}\frac{u^n{\cal{E}}^n([n],\triangle E)}{[n]!}&=&
\sum_{n=0}^{\infty} u^n P_q^{(P)}([n],\triangle E), \nonumber 
\end{eqnarray}
so that the q-Poisson distribution $P_q^{(P)}([n],\triangle E)$ is given by 
\begin{eqnarray}
P_q^{(P)}([n],\triangle E)&=&e_{\frac{1}{q}}^{-{\cal{E}}([n],\triangle E)}\ \frac{{\cal{E}}^n([n],\triangle E)}{[n]!}. 
\end{eqnarray}    

Substituting (44) in (42),
\begin{eqnarray}
\int_{E_1}\cdots \int_{E_k}f^{(q)}_k(E_1,\cdots E_k)dE_1\cdots dE_k&=&\sum_{n-k=0}^{\infty}\frac{[n]!}{[n-k]!}
e_{\frac{1}{q}}^{-{\cal{E}}([n],\triangle E)}\frac{{\cal{E}}^n([n],\triangle E)}{[n]!}, \nonumber \\
 &=&\sum_{m=0}^{\infty} e_{\frac{1}{q}}^{-{\cal{E}}([n],\triangle E)}\ \frac{{\cal{E}}^{m+k}([n],\triangle E)}{[m]!}, \nonumber \\
 &=& {\cal{E}}^k([n],\triangle E).
\end{eqnarray}
From (23), for $r=1$ and using ${\cal{C}}^1_1=1$,
\begin{eqnarray}
{\cal{E}}\{[n],\triangle E\} &=& \int_{E_{\ell}}^{E_u} f^{(q)}_1(E)dE.
\end{eqnarray}
Thus,
\begin{eqnarray}
\int_{E_1}\cdots \int_{E_k}f^{(q)}_k(E_1,\cdots E_k)dE_1\cdots dE_k&=& {\Big( \int_{E_{\ell}}^{E_u}f^{(q)}_1(E)dE\Big)}^k.
\end{eqnarray}
Using (47) in (23), we get for q-Poisson distribution
\begin{eqnarray}
{\cal{E}}\{[n]^r,\triangle E\}&=&\sum_{s=1}^r {\cal{C}}^r_s {\Big( \int_{E_{\ell}}^{E_u}f^{(q)}_1(E)dE\Big)}^s.
\end{eqnarray}

Next, from the generating function (43) of the q-Poisson distribution, it follows 
\begin{eqnarray}
{\cal{G}}^{(s)}(u,\triangle E){\mid}_{u=1}&=&{\Big( {\cal{E}}([n],\triangle E)\Big)}^s \nonumber \\
 &=& {\Big( \int_{E_{\ell}}^{E_u} f^{(q)}_1(E)dE\Big)}^s, 
\end{eqnarray}
so that (48) becomes
\begin{eqnarray}
{\cal{E}}\{[n]^r,\triangle E\} &=& \sum_{s=1}^r {\cal{C}}^r_s {\Bigl\{ {\cal{G}}^{(s)}(u,\triangle E)\Bigr\}}_{u=1}.
\end{eqnarray}
By comparing (41) with (50), 
\begin{eqnarray}
{\cal{C}}^r_s &\equiv & {\cal{B}}^r_s.
\end{eqnarray}
As ${\cal{C}}^r_s$ coefficients do not depend on $f^{q)}_s$, (51) holdsgood in general. Now comparing (23) and (41) and using the
above result (51), {\it{(37) is proved.}}  

\vspace{0.5cm}

The formula (35) for q-Bell number will now be derived using the results of q-Poisson
distribution. The general result (23) for the specific case of q-Poisson distribution upon
using (45) becomes
\begin{eqnarray}
{\cal{E}}\{ [n]^r_{\triangle E}\} &=& \sum_{s=1}^{r} {\cal{C}}^r_s {\Big(
{\cal{E}}\{[n]_{\triangle E}\}\Big)}^s.
\end{eqnarray}
Using the q-Poisson distribution (44) as
\begin{eqnarray}
P_q(n,\lambda)&=& e_{\frac{1}{q}}^{-\lambda}\frac{{\lambda}^n}{[n]!},
\end{eqnarray}
we have 
\begin{eqnarray}
{\cal{E}}\{[N]_{\triangle E}\} &=& \sum_{N=0}^{\infty} [N] P_q(N,\lambda)\ =\ \lambda,
\end{eqnarray}
and 
\begin{eqnarray}
{\cal{E}}\{[N]^r_{\triangle E}\}&=& \sum_{N=0}^{\infty} [N]^r P_q(N,\lambda).
\end{eqnarray}
Then, setting $\lambda = 1$, (52) leads to (35).
  
{\noindent{\bf{VI. q-Product Densities and q-Janossy Function}}}

\vspace{0.5cm}

Janossy [5,6] in his study of a mathematical model for nuclear cascades, introduced a function, which we shall call 'Janossy
Function'. This function is closely related to the 'Product Density Function' of Ramakrishnan  who established [13] a
complete correspondence between the two functions. {\it{The crucial difference between the two functions is contained in their
very definitions}}. Namely, if there are $N$ particles, then $f_h(E_1,E_2,\cdots ,E_h)$ is a product density function of
degree $h$ and $f_h(E_1,E_2,\cdots ,E_h)dE_1dE_2\cdots dE_h$ is the joint probability that there is a particle in the range
$dE_1$, a particle in $dE_2$, etc and a particle in $dE_h$, {\it{irrespective of the number of particles in the other ranges
$dE_{h+1}, dE_{h+2}, \cdots ,dE_N$.}} On the other hand, the Janossy function is ${\psi}_h(E_1,E_2,\cdots , E_h)$ and
${\psi}_h(E_1,E_2,\cdots ,E_h)dE_1dE_2\cdots dE_h$ is the probability that there is a particle in $dE_1$, one in $dE_2$ etc
and one in $dE_h$ and {\it{none in other energy ranges, $dE_{h+1}, dE_{h+2},\cdots ,dE_N$.}} These two functions are related
by 
\begin{eqnarray}
&f_h(E_1,E_2,\cdots ,E_h)={\sum}_{N=h+1}^{\infty} \frac{1}{(N-h)!}& \nonumber \\
 &  \int_{E_{h+1}}\cdots \int_{E_{N}} {\psi}_N(E_1,E_2,\cdots ,E_N)dE_{h+1}dE_{h+2}\cdots dE_N,& 
\end{eqnarray}   
a result derived in [13].

\vspace{0.5cm}

We have introduced q-product density functions in (22), for $[N]$ particles distributed in $dE_1, dE_2, \cdots dE_N$.
Integrating the third relation in (22) over the entire range of $dE_{h+1}dE_{h+2}\cdots dE_N$, and using (21), we obtain 
\begin{eqnarray}
&\Big( \int_{E_{h+1}}\cdots \int_{E_N}f^q_N(E_1,\cdots ,E_N)dE_{h+1}dE_{h+2}\cdots dE_N\Big) dE_1dE_2\cdots dE_h =[N]!& 
\nonumber \\
&  f^{(q)0}_1(E_1)f^{(q)0}_1(E_2)\cdots f^{(q)0}_1(E_h)dE_1dE_2\cdots dE_h. & 
\end{eqnarray}
For the product $f^{(q)0}_1(E_1)f^{(q)0}_1(E_2)\cdots f^{(q)0}_1(E_h)dE_1dE_2\cdots dE_h$, we use the second relation in
(22), so that (57) yields,
\begin{eqnarray}
&f^q_h(E_1,E_2,\cdots ,E_h)=\frac{1}{[N-h]!} & \nonumber \\
& \int_{E_{h+1}}\int_{E_{h+2}} \cdots \int_{E_N}f^q_N(E_1,E_2,\cdots E_N)dE_{h+1}dE_{h+2}\cdots dE_N. & 
\end{eqnarray}

\vspace{0.5cm}

We now introduce q-Janossy function ${\psi}^q_h(E_1,E_2,\cdots E_h)$ such that\\  ${\psi}^q_h(E_1,E_2,\cdots
,E_h)dE_1dE_2\cdots dE_h$ represents the probability that there is $[h]/h$ particle in $dE_1$, $[h]/h$ in $dE_2$ etc and
$[h]/h$ in $dE_h$ and none in other energy ranges. It is to be noted here that q-Janossy function is distinctly different
from the ordinary Janossy function and only when $q=1$ they coincide. To relate q-Janossy function with the q-product
density function, we first write
\begin{eqnarray}
{\psi}^q_h(E_1,E_2,\cdots ,E_h)&=& P^q(h)f^q_h(E_1,E_2,\cdots ,E_h),
\end{eqnarray}
where the factor $P^q(h)$ ensures that there are $[h]$ particles in the entire energy range. By integrating (59) over the
entire range of $dE_1dE_2 \cdots dE_h$ and using (57) with $N=h$, we find
\begin{eqnarray}
P^q(h)&=&\frac{1}{[h]!}\int_{E_{1}}\int_{E_{2}}\cdots \int_{E_{h}}{\psi}^q_h(E_1,E_2,\cdots ,E_h)dE_1dE_2\cdots dE_h.
\end{eqnarray}
Now, $f^q_h(E_1,E_2,\cdots ,E_h)$ with $N=h$ can be related to $f^q_h(E_1,E_2,\cdots ,E_h)$ for $N$, by a weighted sum of
the later over $h+1$ to $\infty$, i.e.,
\begin{eqnarray}
f^q_h(E_1,E_2,\cdots ,E_h){\mid}_{[N]=[h]}&=&\sum_{N=h+1}^{\infty} P^q(N)f^q_h(E_1,E_2,\cdots ,E_h),
\end{eqnarray}
where the $f^q_h$ in the right side of (61) stands for the q-product density of degree $h$ {\it{when $[N]$ number of
particles are present}}. Upon identifying this with (58) and then using (59) with $h=N$, we obtain 
\begin{eqnarray}
&f^q_h(E_1,E_2,\cdots ,E_h){\mid}_{[N]=[h]}= \sum_{N=h+1}^{\infty} \frac{1}{[N-h]!}& \nonumber \\
&  \int_{E_{h+1}}\int_{E_{h+2}}\cdots \int_{E_{N}} {\psi}^q_N(E_1,E_2,\cdots ,E_N)dE_{h+1}\cdots dE_N, & 
\end{eqnarray}
a relation connecting q-Janossy function and q-product density function.            

\vspace{0.5cm} 

{\noindent{\bf{V.Discussion and Summary}}}

\vspace{0.5cm}

We have extended the theory of product densities to q-product densities by considering the number of particles
$[n]$ (a q-number) as a stochastic variable. This consideration at this stage is a mathematical possibility. It is also
conceivable to think of q-number of particles distributed either by scaling the energy intervals $dE_1,dE_2,\cdots ,dE_N$
and restricting to a subset of the intervals or by addressing a situation in which the measurement of the number of
particles has uncertainties. This stochastic variable takes on  
 discrete values (for a given $q$) over a continuous variable $E$. $E$ is
treated as ordinary variable and so the integrals involving $E$ are ordinary integrals. Such a q-point process can be consistently
described by q-product densities which are introduced for the first time here. The $r^{th}$ moment of $[n]$ is described in terms
of q-generalized ${\cal{C}}^r_s$ coefficients with falling factorials. For a fixed $[N]$, ${\cal{C}}^r_s$ coefficients are
identified with q-Stirling numbers of the second kind. A recursion relation for ${\cal{C}}^r_s$ is obtained. This leads to $q$-Bell
number and $q$-Dobinsky formula which is derived here using only the q-product densities and this agrees with the result obtained
using q-boson coherent states. The notion of q-product densities is further investigated using q-probability distribution and its
generating function. The case of q-Poisson distribution and its generating function is discussed in detail. 

\vspace{0.5cm}

A q-generalization of Janossy function is given and its relation with the q-product density function is derived. 

\vspace{0.5cm}

In this approach, since the continuous variable $E$ is taken to be the ordinary variable, the calculus of differentiation
with respect to and integration over $E$ remain unaltered. Consequently the evolution equations for the q-product densities
will be very similar to those of the usual product densities.        

\vspace{0.5cm}

{\noindent{\bf{Acknowledgement}}}

\vspace{0.5cm}

Useful discussions with S.K.Srinivasan which triggered this study are acknowledged with thanks.

\vspace{0.5cm}

{\noindent{\bf{References}}}

\vspace{0.5cm}

\begin{enumerate}
\item M.S.Bartlett. {\it{Stochastic Processes}} (Cambridge University Press, 1955).
\item H.J.Bhabha and W.Heitler. {\it{The passage of fast electrons and the theory of cosmic
showers}} Proc.Roy.Soc. (London) {\bf{A159}} (1937) 432-458.
\item A.T.Barucha-Reid. {\it{Elements of the theory of Markov Process and their Applications}}
(McGraw-Hill, 1960).
\item H.Exton. {\it{q-Hypergeometric Functions and Applications}} (Ellis Harwood Ltd, 1983).   
\item L.Janossy. {\it{On the absorption of a nucleon cascade}} Proc.Royal Irish Academy
{\bf{A53}} (1950) 181-188.
\item L.Janossy. {\it{Cosmic Rays}} (Oxford University Press, 1950).
\item J.Katriel. {\it{Bell Numbers and Coherent States}} Phys.Lett. {\bf{A273}} (2000) 159-161.
\item J.R.Klauder and E.C.G.Sudarshan. {\it{Fundamentals of Quantum Optics}} (W.A.Benjamin Inc,
1968).
\item S.C.Milne. {\it{A q-analogue of restricted growth functions, Dobinski's equality, and
Charlier polynomials}} Trans.Amer.Math.Soc. {\bf{245}} (1978) 89-117.
\item A.M.Perelemov. {\it{On the completeness of some subsystems of q-deformed coherent
states}} Helv.Physica.Acta {\bf{68}} (1996) 554-576.
\item J.Pitman. {\it{Some probabilistic aspects of set partitions}} Amer.Math.Monthly
{\bf{104}} (1997) 201-208.
\item A.Ramakrishnan. {\it{Stochastic Processes relating to particles distributed in a
continuous infinity of states}} Proc.Camb.Phil.Soc. {\bf{46}} (1950) 595-602.
\item A.Ramakrishnan. {\it{A note on Janossy's mathematical model of nucleon cascade}} Proc.Camb.Phil.Soc {\bf{48}} (1952) 451-456.
\item A.Ramakrishnan. {\it{Probability and Stochastic Processes}} in Handbuch der Physik. Band
III/2 (Springer, 1959) 524-651.
\item S.K.Srinivasan. {\it{Stochastic Theory and Cascade Processes}} (Elsevier Pub., 1969).       
\end{enumerate} 
\end{document}